\documentclass{article}

\usepackage{microtype}
\usepackage{graphicx}
\usepackage{subfigure}
\usepackage{booktabs} 

\usepackage{hyperref}

\usepackage[accepted]{icml2024}
\usepackage{amsmath}
\usepackage{amssymb}
\usepackage{mathtools}
\usepackage{amsthm}

\usepackage{microtype}
\usepackage{graphicx}
\usepackage{subfigure}
\usepackage{multicol}
\usepackage{svg}
\usepackage{booktabs} 
\usepackage{multirow}
\usepackage{comment}

\usepackage[utf8]{inputenc} 
\usepackage[T1]{fontenc}    
\usepackage{hyperref}       
\usepackage{url}            
\usepackage{booktabs}       
\usepackage{amsfonts}       
\usepackage{nicefrac}       
\usepackage{microtype}      
\usepackage{xcolor}         

\usepackage{amsmath}
\usepackage{amssymb}
\usepackage{mathtools}
\usepackage{amsthm}

\usepackage[capitalize,noabbrev]{cleveref}

\theoremstyle{plain}

\theoremstyle{definition}

\theoremstyle{remark}

\usepackage[textsize=tiny]{todonotes}

\newcommand{\enc}[0]{\operatorname{enc}}
\newcommand{\dec}[0]{\operatorname{dec}}
\newcommand{\quant}[0]{\operatorname{quant}}
\newcommand{\bz}[0]{\boldsymbol{z}}

\icmltitlerunning{Spiking Music: Audio Compression with Event Based Auto-encoders}

\begin{document}

\twocolumn[
\icmltitle{Spiking Music: Audio Compression with Event Based Auto-encoders}

\begin{icmlauthorlist}
\icmlauthor{Martim Lisboa}{EPFL}
\icmlauthor{Guillaume Bellec}{EPFL}
\end{icmlauthorlist}

\icmlaffiliation{EPFL}{Laboratory of Computational Neuroscience, EPFL, Lausanne}

\icmlcorrespondingauthor{Guillaume Bellec}{guillaume.bellec@epfl.ch}

\icmlkeywords{Machine Learning, ICML}

\vskip 0.3in
]



\footnotetext[1]{EPFL, address communications to guillaume.bellec@epfl.ch}
\let\thefootnote\relax\footnotetext{This is a pre-print article.}


\begin{abstract}
Neurons in the brain communicate information via punctual events called spikes. The timing of spikes is thought to carry rich information, but it is not clear how to leverage this in digital systems. We demonstrate that event-based encoding is efficient for audio compression. To build this event-based representation we use a deep binary auto-encoder, and under high sparsity pressure, the model enters a regime where the binary event matrix is stored more efficiently with sparse matrix storage algorithms. We test this on the large MAESTRO dataset of piano recordings against vector quantized auto-encoders. Not only does our "Spiking Music compression" algorithm achieve a competitive compression/reconstruction trade-off, but selectivity and synchrony between encoded events and piano key strikes emerge without supervision in the sparse regime.
\end{abstract}

\section{Introduction}

In most organisms like insects or primates, neurons exchange information via short electrical pulses called spikes. Biological neurons encode information in their spike-timing or their spike-count and send information collectively across brain areas in the form of complex spike packets.
There exist many theories about the potential computational benefits of spikes, on top of their necessity for communication over long-nerve axons, a recurring idea is that the brain has to minimize the number of spikes to reduce the energy it consumes and leverage parsimonious representations \cite{olshausen1996emergence,tkavcik2010optimal,chalk2018toward,ocko2018emergence,padamsey2022neocortex}. 

In the digital world, event-based encoding allocates more bits for complex events and change points to avoid redundancy.  
This principle is for instance used in classical MPEG video codecs which are engineered to rely on change points and predictions.
In the era of deep learning, we seek for an algorithm that can learn event-based representations of conceptual change points and physical events at multiple levels of abstraction.
As of today, the most promising deep neural compression algorithms do not use such an event-based representation, rather they use vector quantized variational autoencoders (VQ-VAE) to encode audio signals into a matrix of integers \cite{van2017neural,garbacea2019low,defossez2022high}: the audio is encoded with a low but rigid bit rate and VQ-VAE cannot be extended easily towards an event-based compression algorithm.

Musical audio recordings are very likely to benefit from event-based representations. There, the timbre of an instrument can be encoded at rare change points; while note onsets could be encoded separately with precisely timed events.
To approximate the potential advantages of event-based encoding for audio compression, we turn our attention to the MAESTRO dataset, which encompasses recorded piano competition performances with both audio recordings and exact MIDI renditions of the pianist's key strike and pedal action events. Recorded MIDI data is sufficiently precise for the performance to be judged on a different MIDI piano in a separate room \cite{hawthorne2018enabling}.
While we expect a compression bit rate of audio signals between $1$ and $10$~kbps (kilobits per second) with VQ-VAEs -- already an order of magnitude smaller than classical compression algorithms \cite{defossez2022high} -- the uncompressed MIDI-files represent only $0.12$~kbps on this dataset. So assuming the remaining information in the audio signal needs to be encoded rarely (e.g. the piano sound font, room reverberation, microphone location), we speculate that event-based compression should be able to reduce the compression bit rate by another order of magnitude. With this goal in mind, we design the \emph{Spiking Music compression} algorithm.


In \emph{Spiking Music compression}, we do not model biological neural networks. Instead, we leverage successful deep auto-encoder architectures to build an event-based auto-encoder. In short, we replace the VQ operator at the bottleneck of the VQ-VAE with a binarized representation. This compressed representation is therefore a binary matrix $\bz \in \{0,1\}^{N,T_z}$ where $N$ is the number of units and $T_z$ is the number of time steps. When the event matrix $\bz$ becomes sparse, we show that the compressed audio can be stored with $\mathcal{O}(\log_2\left(T_z\right))$ instead of the linear scaling $\mathcal{O}\left(T_z\right)$ with VQ models. We show on the MAESTRO dataset that Spiking Music compression is indeed viable in both the dense and sparse regimes. In this first attempt, we achieve a competitive audio reconstruction quality on par with VQ models at $3$kbps. Analyzing the relationship between the event matrix that is learned without supervision and the note onsets given in the MIDI files, we discover an emergent feature in the sparse regime: the binary units become synchronized and selective to specific note onsets. This confirms the intuition that \emph{Spike Music compression} can discover a mapping between encoded events and physical events.


\paragraph{Related work}

The representation of audio signals into a sequence of integers with VQ-VAE is becoming a promising approach for audio compression \cite{defossez2022high,garbacea2019low}.
Whether the decoder is trained via adversarial loss functions \cite{defossez2022high} or diffusion-based objectives \cite{sohl2015deep,ho2020denoising,preechakul2022diffusion}, the compression bit rate is dictated by the nature of VQ layer placed after the encoder. 
We are only aware of one recent paper that attempted to build an event-like representation \cite{dieleman2021variable} by encouraging slowness and therefore repetition in the latent code before quantization. 
Here we employ a different strategy that gets rid of quantization strategies altogether and uses instead a simpler end-to-end binarized neural network \cite{bengio2013estimating,courbariaux2016binarized,Esser2016}.
While there is a long history of binary auto-encoders starting with deep belief networks \cite{deng2010binary,song2018self}, binary autoencoders have never been exploited for as compression as far as we know. More crucially, we are not aware of any paper that has explored binary auto-encoders in the regime where they are stored more economically as a sparse matrix.
It might appear that by generating information as a sparse matrix, our algorithm loses compatibility with audio generative autoregressive models~\cite{dhariwal2020jukebox,zeghidour2021soundstream,copet2023simple} which operates with integer matrices: however, any $n$-bit binary vector can be equivalently represented as a $n$-bit integer. So the same technique applies without limitations, and additionally, sparse event matrices can benefit from the run-length transformers \cite{robinson1967results,dieleman2018challenge} to avoid successive generations of null tokens.

In the fields of computational neuroscience and neuromorphic computing, it was speculated three decades ago that spike-timing \cite{richmond1987temporal,thorpe1996speed,maass1995computational,gerstner2014neuronal} can encode rich information, but we are not aware of a deep-learning event-based model for compression. The conceptual relationship between deep binary networks and spiking models has enabled the simulation of biological neurons with deep learning frameworks \cite{neftci2019surrogate,bellec2021fitting,sourmpis2023trial}. These bio-inspired model was used sometimes for speech transcription \cite{bellec2020solution,cramer2020heidelberg,pan2020efficient}, musical melody \cite{liang2021stylistic}, recognize instrument \cite{shah2022utilizing}, speech denoising \cite{timcheck2023intel}, but not audio compression.
While our analysis and model operate in discrete time, we expect that the event-timing conveys more information with smaller timesteps or continuous time representation. With the encouraging progress of optimization in continuous time models \cite{comcsa2021temporal,comcsa2021spiking,wunderlich2021event,kajino2021differentiable,subramoney2022efficient,azabou2023unified,stanojevic2023exact,stanojevic2023training}, we believe that there might be more opportunities to study event-based compression in continuous time in the future.


In summary, our contributions to the field of neural audio compression are: 
(i) To show that simple binary auto-encoders are performing surprisingly well on neural audio compression benchmarks. (ii) We exhibit functional sparse binary auto-encoders operating in the sparse regime. (iii) We observe the emergence of an event-based coding that is selective and synchronized with piano key-stroke which demonstrates that it encodes high-level features.
For the field of neuromorphic computing, spiking neurons have been penalized on standard machine learning benchmarks by the difficulty of optimization in event-based models. By opening the way in event-based audio compression, we aim to set an inspiring benchmark in which event-based models are likely to finally leverage a computational advantage.

\label{background}

\section{An event-based neural compression method}

\begin{figure*}
    \centering
    \includegraphics[width=0.9\textwidth]{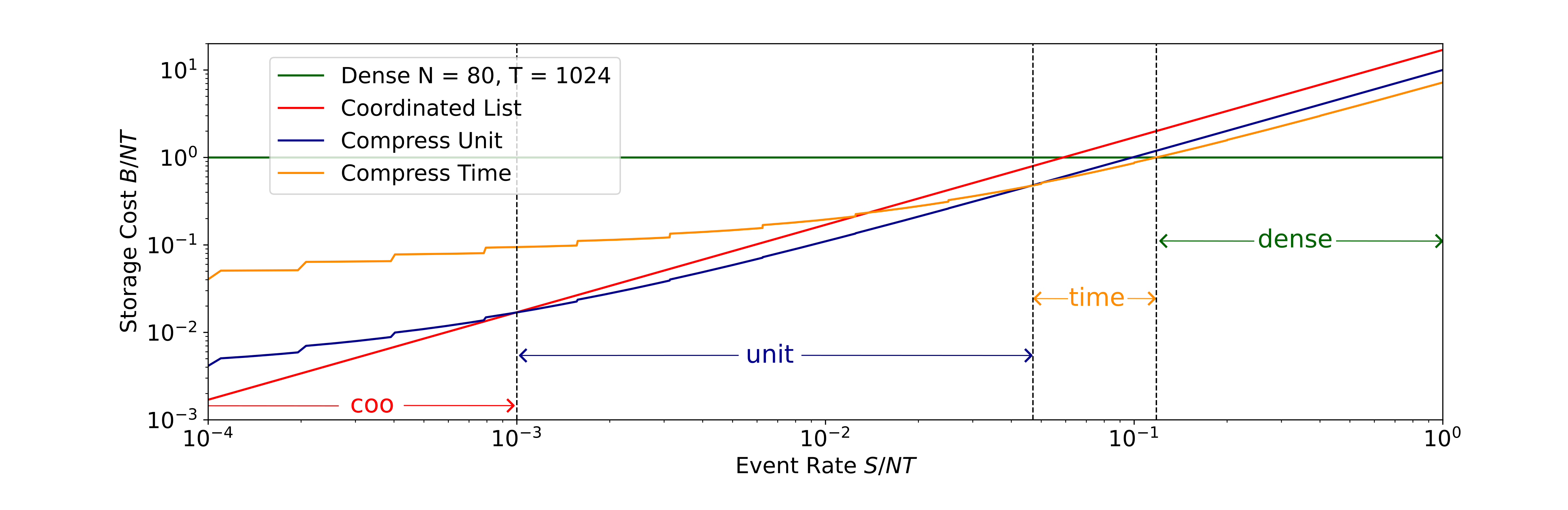}
    \caption{\textbf{Storage of sparse binary matrices} Given a binary event matrix $\boldsymbol{z}$ with $N=80$ units, $T=1024$ time steps, and $S$ events, the storage cost is given by an exact formula. There are four regimes where each of the 4 matrix storage formats is optimal.}
    \label{fig:sparsity}
\end{figure*}

\paragraph{Recall on neural compression}
Contemporary neural compression methods \cite{balle2016end,van2017neural,garbacea2019low,defossez2022high,zeghidour2021soundstream} encode an audio waveform $\boldsymbol{x} \in \mathbb{R}^{T_x}$ into a compressed vector quantized representation denoted by $\bz_{VQ} \in \{0, \cdots K\}^{Q \times T_z}$ ($Q$ is the number of code-books and $K$ is the size of each code-book size).
It relies on a deep encoder $\operatorname{enc}$, a deep decoder $\operatorname{dec}$ and a quantization operator $\operatorname{quant}$ (e.g. vector quantization "VQ" or residual vector quantization "RVQ") such that the input signal $\boldsymbol{x}$ is reconstructed as $\widehat{\boldsymbol{x}} = \dec \left( \bz_{VQ} \right)$ with $\bz_{VQ} = \quant \left( \enc (\boldsymbol{x}) \right)$. The training algorithm minimizes a linear combination of the reconstruction loss $\mathcal{L}_x$ and an auxiliary representation loss $\mathcal{L}_z$ (see \cite{van2017neural,defossez2022high}). In recent works, the reconstruction loss $\mathcal{L}_x$ often involves a diffusion model or GAN discriminators. Then the storage cost achieved with neural compression is quantified by the number of bits $B_{\mathrm{VQ}}$ that is required to store the quantized representation $\boldsymbol{z}$:
\begin{equation}
B_{\mathrm{VQ}} = Q T_z \lceil\log_2\left(K\right)\rceil.
\label{eq:B_VQ}
\end{equation}
It is also possible to compress this discrete representation further using entropy code \cite{balle2016end} or \cite{defossez2022high}. For instance, leveraging Huffman encoding and language models can reduce the code length of each integer\cite{deletang2023language}, but it would still require at least one bit per time step, so the storage cost $B_{\mathrm{VQ}}$ remains $\mathcal{O}(T_z)$. By post-processing the integer sequence with run-length encoding \cite{robinson1967results}, one can theoretically reduce it sub-linearly, but integers generated by VQ-VAE appear to vary too frequently to benefit from run-length encoding natively \cite{dieleman2021variable}.
In practice, training a transformer \cite{defossez2022high} to leverage model-based entropy code is costly and reduces the bit rate by roughly $30\%$ with VQ-VAE.
Although not negligible, this is much less than the expected difference between classical and neural compression algorithms.
So we compare directly the storage cost $B$ for the quantized/binary representation of auto-encoders and keep in mind that storage costs with less than $30\%$ difference can be compensated with post-processing.

\paragraph{Binary sparse matrix storage} 
In Spiking Music compression, the compression bit rate is given by the storage cost $B$ of the latent representation in an auto-encoder architecture.
For a binary matrix of size $\boldsymbol{z} \in \{0,1\}^{N\times T_z}$ there are multiple ways to store this matrix. We consider in particular standard sparse matrix representation formats \cite{eisenstat1977yale}.  Below we detail the computation of the storage cost $B$ of this matrix as a function of $N$, $T_z$, and the number of events: $S = \sum_{t,i} z^i_t$. To illustrate the calculation of $B$ we consider the following example with $N=5$ units, $T=10$ time-steps, and $S = 7$ events:
\begin{equation}
    \boldsymbol{z} = \begin{bmatrix}
        0 & 0 & 0 & 1 & 0 & 0 & 0 & 0 & 0 & 0 \\
        1 & 0 & 0 & 0 & 1 & 0 & 0 & 0 & 0 & 0 \\
        0 & 0 & 0 & 0 & 0 & 0 & 0 & 0 & 0 & 1 \\
        0 & 1 & 0 & 1 & 0 & 0 & 0 & 0 & 0 & 0 \\
        0 & 0 & 0 & 0 & 0 & 0 & 1 & 0 & 0 & 0 \\
    \end{bmatrix}
    \label{eq:z_example}
\end{equation}
The dense storage format views  $\boldsymbol{z}$ as a dense matrix of binary numbers. It yields a storage cost $B_{\mathrm{dense}} = NT_z$~bits which increases linearly with $T_z$. In the example above: $B_{\mathrm{dense}}=50$bits.

Without loss of information, the matrix $\boldsymbol{z}$ above can be stored as a list of unit and time indices: $M_{\mathrm{coo}} = [(0,3);(1,0);(1,4);(2,9);(3,1);(3,3);(4,6)]$, so assuming the indices are respectively encoded with $\lceil\log_2\left(N\right)\rceil$ and $\lceil\log_2\left(T\right)\rceil$~bits, we find:
\begin{eqnarray}
    B_{\mathrm{coo}}
    & = & S \left( \lceil\log_2\left(N\right)\rceil + \lceil\log_2\left(T\right)\rceil \right) \\
    & = & 7 \times \left(3+4\right) = 49 \text{bits}~. \nonumber
\end{eqnarray}
Note that in comparison with the standard coordinated list sparse matrix format (coo), here it is not needed to store the value of the non-zero entries since it's always $1$. 

\paragraph{Compressed time/units formats} While the coordinated list format already achieves the desired complexity $\lceil\log_2\left(T\right)\rceil$, it is not always optimal. One can avoid repeating the unit index $i$ by storing the lists of time events of each unit in a specific order, for instance in the example above $[(3);(0,4);(9);(1,3);(6)]$). This corresponds to a binary extension of compressed-row sparse matrix format \cite{eisenstat1977yale} where the value of the non-zero entries is spared. Following the event-encoding analogy we refer to this format as the \emph{compressed time} format. Under this format the matrix is stored with two lists: the first list $L_\mathrm{time}$ contains the concatenation of the time indices for all units, and the second $L_\mathrm{units}$ has length $N+1$ and gives the indices in $L_\mathrm{time}$ at which the unit index changes (i.e. the list of time events of unit $i$ are $L_{\mathrm{time}}[L_{\mathrm{units}}[i]: L_{\mathrm{units}}[i+1]]$. For the example above, we can reconstruct $\bz$ from the two lists:
\begin{equation}
    M_{\mathrm{time}} = \begin{cases}
        L_{\mathrm{time}} &= [3,0,4,9,1,3,6] \\
        L_{\mathrm{units}} & = [0,1,3,4,6,7]
    \end{cases}
\end{equation}
To compute the storage cost $B_{\mathrm{row}}$ for this compressed row format, we observe that $L_{\mathrm{time}}$ contains $S$ integers of $\lceil\log_2\left(T\right)\rceil$ bits and $L_{\mathrm{units}}$ contains $N+1$ integers smaller than $S+1$. Since the first is always $0$ and the last is always $S$ and they do not need to be stored. So the amount of bits required to store $\boldsymbol{z}$ with the format is:
\begin{eqnarray}
    B_{\mathrm{time}} & = & S \lceil\log_2\left(T\right)\rceil + \left(N-1\right) \lceil\log_2\left(S\right)\rceil \\
    & = & 7\times4 + 4\times3 = 40\text{bits}~. \nonumber
    \label{eq:B_ctime}
\end{eqnarray}
As for $B_{\mathrm{coo}}$, the cost $B_{\mathrm{time}}$ is dominated by $S \lceil\log_2\left(T\right)\rceil$ and it may be lower than $B_{\mathrm{coo}}$ depending on the values of $N$ and $S$ (see Figure \ref{fig:sparsity}).
Symmetrically, we can inverse the roles of rows and columns (i.e. time and unit indices) leading to the compressed column format with the storage cost:
$B_{\mathrm{units}} = S \lceil\log_2\left(N\right)\rceil + \left(T-1\right) \lceil\log_2\left(S\right)\rceil=7\times3 + 9 \times 3 =48 $bits.

\begin{figure*}[ht]
    \centering
    \includegraphics[width = 0.8\textwidth]{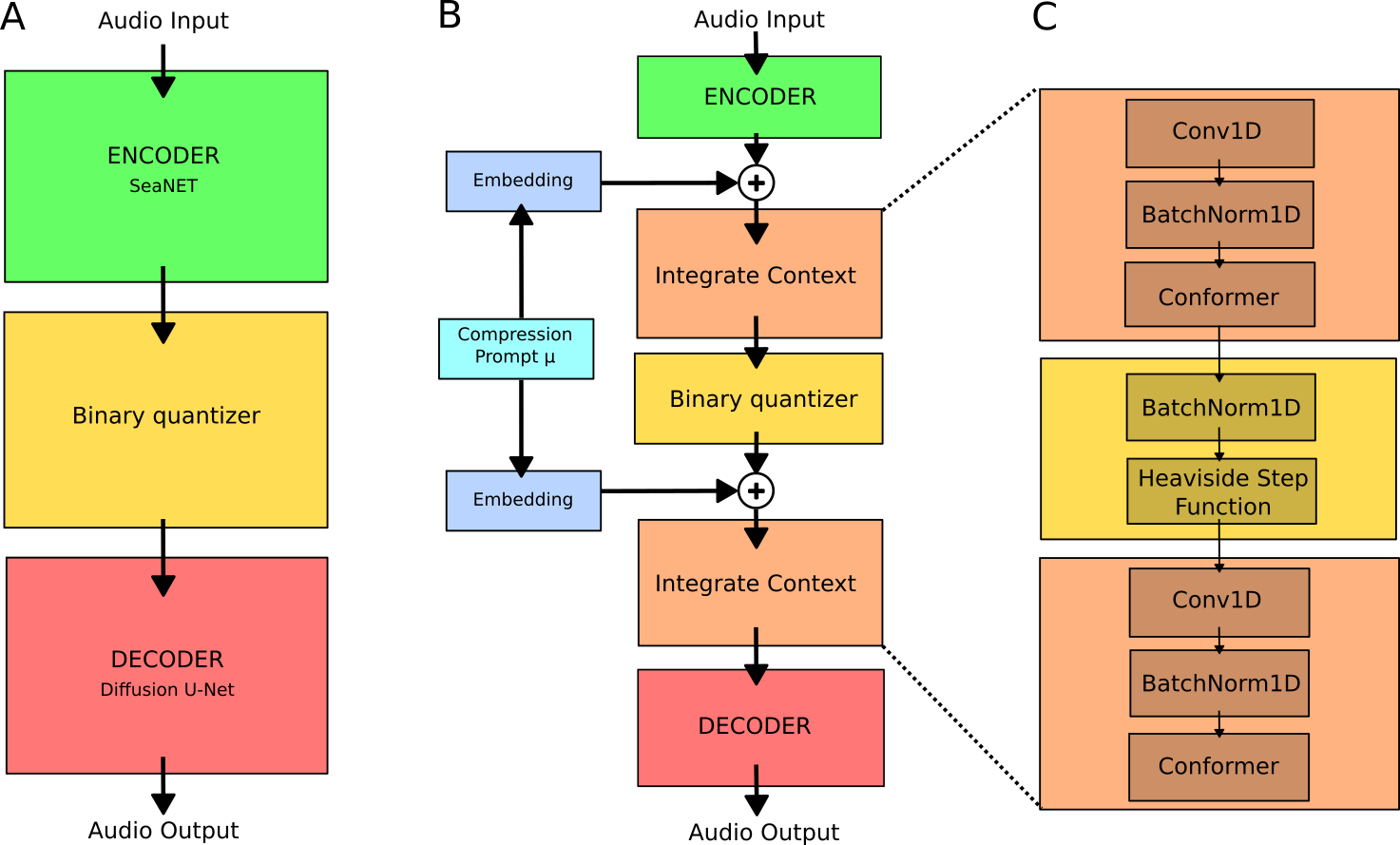}
    \caption{\textbf{A binary auto-encoder architecture.} \textbf{A} Our encoder is taken from \cite{tagliasacchi2020seanet,defossez2022high} and we report results using the Mousai diffusion decoder \cite{preechakul2022diffusion}. \textbf{B} Architecture of the SPARSE model. The embeddings and compression prompt $\mu$ are only used in the $\mu$-SPARSE model. \textbf{C} Details of the "Integrate context" and "Binary quantizer" blocks.}
    \label{fig:model_simple}
\end{figure*}

\paragraph{Sparse and dense regimes} In Figure \ref{fig:sparsity} we display for each sparse matrix format how the different storage costs $B$ as a function of the number of events $S=\sum_{t,i} z^i_t$.
Remarkably, there are four distinct regimes in which each of the four matrix storage formats is optimal.
In the following, we target the sparse regime where the compressed time format is optimal. We can however only target a sparsity level "on average" and individual samples may be optimally stored with the dense format or another sparse format. In practice, one can offer this flexibility by adding a negligible overhead of $2$~bits to indicate which format is used for each sample. There is also an additional overhead of $\lceil\log_2\left(S_{\mathrm{max}}+1\right)\rceil$bits per sample, where $S_{\mathrm{max}}$ correspond to the audio sample with the largest number of spikes in the dataset. This is because we need to store the length of the coordinate lists.

\paragraph{Spiking Music compression: training and architecture}
We will now describe the auto-encoder architecture for Spiking Music compression and how to train it.
We describe two variants, the FREE and the SPARSE Spiking music compression models, it correspond to two distinct regimes where $\boldsymbol{z}$ is optimally encoded as a dense matrix where the optimal storage cost is $B_{\mathrm{dense}} = N T_z$ or a sparse matrix format where the optimal storage cost is $B_{\mathrm{sparse}} = \mathcal{O}(S \lceil\log_2\left(T_z\right)\rceil$ (this is typically the compressed time format in our case). The purpose of the FREE model is to test Spiking Music compression in the dense regime where the binary encoder acts similarly as a vector quantized auto-encoder. The SPARSE model reflects best the analogy of event-based encoding.

In both cases, we build an auto-encoder that is end-to-end differentiable and defines the latent representation as a binary matrix. We keep the same encoder $\mathrm{enc}$, decoder $\mathrm{dec}$, and reconstruction loss $\mathcal{L}_x$ unchanged from our VQ-VAE re-implementation, but we replace the $\quant$ operator with a binary quantizer as follows:
\begin{equation}
\boldsymbol{z} = H \circ BN \circ L \circ \enc(\boldsymbol{x}),
\label{eq:H_L_enc_x}
\end{equation}
In the simpler form, $L$ is a linear layer, $BN$ is a batch norm layer, and $H$ is a Heaviside function with a surrogate gradient to optimize $\mathrm{enc}$ and $\mathrm{dec}$ via back-propagation end-to-end \cite{bengio2013estimating,Esser2016,neftci2019surrogate,bellec2018long}. Concretely, we assign the pseudo derivative $H'(l)=~\max(0, 1 - |l|)$ to the Heaviside function $H$ to enable auto-differentiation throughout the computational graph. Without auxiliary loss functions, this defines the FREE Spiking Music compression model. In summary, it is a simple replacement of the RVQ model with a normalized linear layer and a step function, but it performs surprisingly well in comparison to our RVQ baseline.

\paragraph{Sparse Spiking Music compression}
To define a SPARSE model that is optimal for the compressed time format, we made two major modifications: First, we added an auxiliary loss function $\mathcal{L}_z$ on the latent representation to enforce sparsity.
This loss $\mathcal{L}_z$ is a function of the number of events $S=\sum_{t,i} z^i_t$, we could simply choose $\mathcal{L}_z=S=\sum_{t,i} z^i_t$ but we opted to set a target compression-rate $B_0$ below which the sparsity loss is clipped. So unless specified otherwise, we define:
\begin{equation}
\mathcal{L}_z = \max(0, B_{\mathrm{sparse}} - B_{0})~.
\label{eq:Lz}
\end{equation}
The optimization minimizes the weighted loss $\mathcal{L} = \mathcal{L}_x + \gamma \mathcal{L}_z$ where $\gamma$ follows a monotonous schedule growing smoothly from $\gamma=0$ to $\gamma_{\infty}$.

The second modification for the SPARSE model anticipates that when the encoder or decoder is convolutional with a limited receptive field, it is not possible to integrate information from a wider context to cancel redundant information. Hence two conformer-based layers are added before the layer $L$ and after $H$ in equation \eqref{eq:H_L_enc_x}, see Figure \ref{fig:model_simple} for details.

\section{Compression of piano recordings}
\label{model}

\paragraph{Dataset and training setup}
To benchmark the different neural compression algorithm
we rely on the MAESTRO dataset \cite{hawthorne2018enabling}. This dataset contains more than 170 hours of virtuosic piano performances, finely aligned with MIDI data describing, notably, note onset and release events. All models were trained using batches of 32 clips from this dataset, with a sampling rate of $22.05$kHz. The duration of the encoded audio clips is $2^{17}$ samples (approximately $6s$). 

All the tested neural compression algorithms rely on a comparable auto-encoder architecture with three distinct parts: encoder, bottleneck, and decoder, as sketched in Figure \ref{fig:model_simple}.  For the encoder, we re-use the SeaNet convolutional architecture from \cite{tagliasacchi2020seanet,defossez2022high}. For the decoder, we explored both using the GAN-trained decoder from \cite{defossez2022high} and the diffusion decoder from \cite{preechakul2022diffusion}, but below we report the results obtained with the Mousai diffusion-based decoder  \cite{preechakul2022diffusion} using  $\upsilon$-objective diffusion \cite{ho2022classifier} and a U-Net \cite{schneider2023mo} architecture with 9 nested blocks of increasing channel counts $[8, 32, 64, 128, 256, 512, 512, 1024, 1024]$ and down-sampling factor $[1, 4, 4, 4, 2, 2, 2, 2, 2]$. All the models are trained for $1$ million training steps.

\begin{figure}[ht]
    \centerline{\includegraphics[width=0.9\columnwidth]{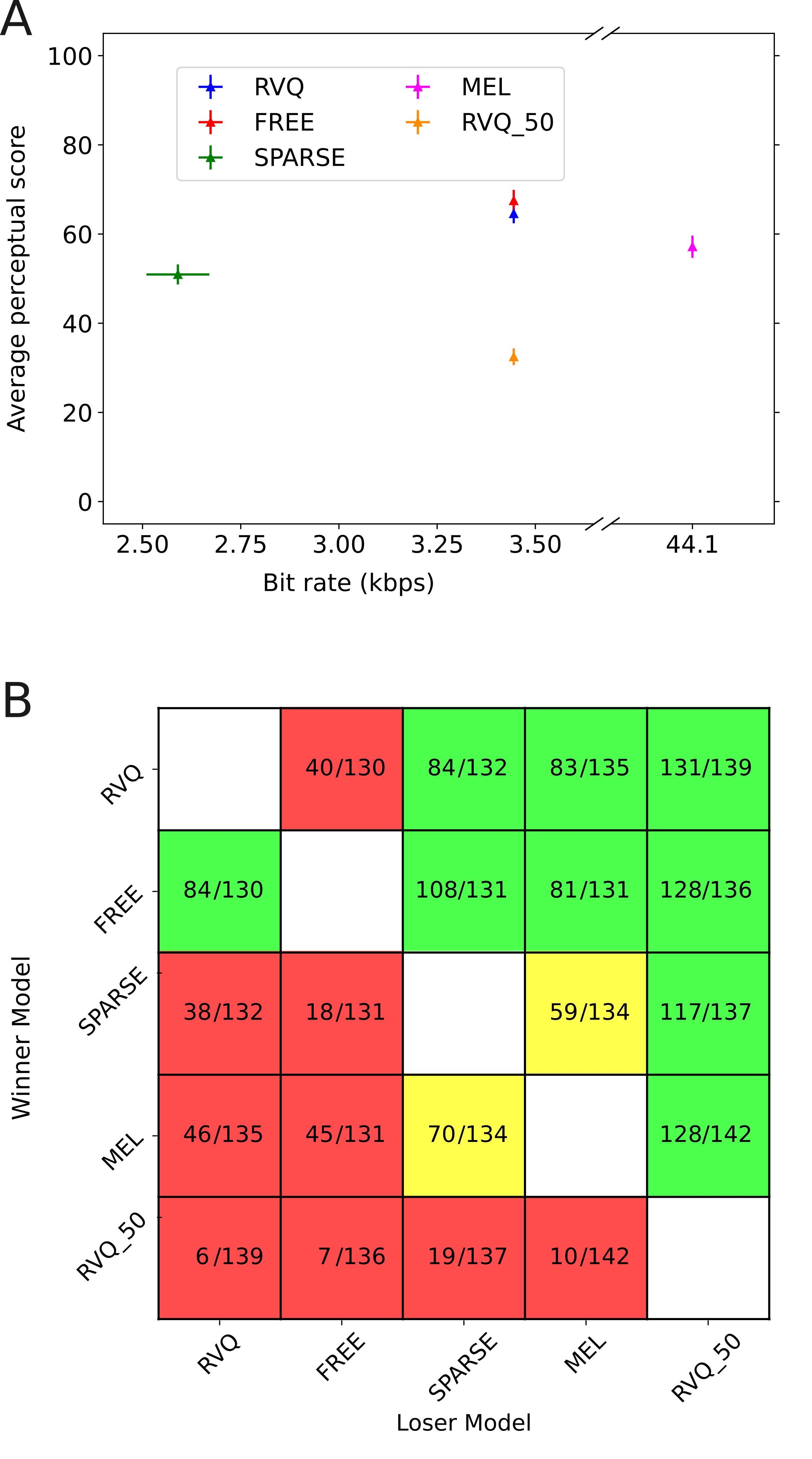}}
    \caption{\textbf{Perceptual quality ratings.} \textbf{A} Summary of the MUSHRA ratings of audio reconstruction quality. \textbf{B} Number of listening tests where the Winner model was rated better than the Loser model. A yellow cell means that the number of wins is not significantly different from the number of losses.}
    \label{fig:exp}
\end{figure}

\paragraph{RVQ baseline}
As far as we know, the current state-of-the-art of neural audio compression \cite{defossez2022high} relies on a vector RVQ bottleneck between the encoder and the decoder. This RVQ bottleneck is a stack of residual vector-quantizers \cite{defossez2022high}.
To estimate the compression rate of this baseline, we count the number of bits necessary to store the dense matrix of integers encoded by the bottleneck. In our setting, equation \eqref{eq:B_VQ} gives $B_{VQ} = 80$ bits/time step, using 8 vector quantizers, each with $2^{10}$ codewords. Using a downsampling factor of 512 in the encoder, this baseline encodes single-channel audio at $22.05$kHz at $3.445$kbps. We targeted this compression rate because the VQ-VAE in \cite{defossez2022high} produced recognizable and realistic audio samples but human listeners with headphones can still perceive differences with the reference wave file.

\paragraph{Comparison with similar bit budget} For Spiking Music Compression, the output of the encoder is designed to yield the same sampling rate as for the RVQ baseline with a dense encoding. This is achieved by setting the number of binary units to $N=80$. This given a storage cost $B_{dense}= 80 \times T_z$ which is the same as for the RVQ baseline and therefore yields the same compression bit-rate: $3.445~$kbps. For the SPARSE model, we set the threshold $B_{0}=B_{dense}$ of sparsity loss function $\mathcal{L}_z$ to the corresponding bit-count and manually tuned $\gamma_\infty$ such that the trained SPARSE model reached a lower compression bit-rate than the dense and RVQ model on average without collapsing. The audio samples are provided as supplementary materials.


\paragraph{SI-SNR and MUSHRA score} To evaluate the quality of the audio reconstruction from each compressed representation, we compute numerically the Scale-Invariant Signal-to-Noise Ratio (SI-SNR) \cite{defossez2022high,luo2019conv,nachmani2018fitting,chazan2021single} and complement this with a perceptual quality score reported by human listeners on a crowd-sourcing platform.
This perceptual metric follows the MUSHRA protocol \cite{bs20151534}. Five $6s$ long audio samples from the MAESTRO dataset were chosen to avoid musical genres and compositions that would be uncommon to the average listener. Participants were asked to rate the resemblance between the audio reference and $5$ compressed audios obtained with the different algorithms. Ratings are given as a score varying between 0 and 100, aligned with a qualitative scale that divides scores into 5 equally spaced categories: "Bad", "Poor", "Fair", "Good" and "Excellent", from low to high.  Among the samples to be rated, there is a hidden copy of the reference as well as an experimental manually distorted anchor. When a participant does not classify the reference as "Excellent" and the anchor as "Bad", the ratings are discarded. 

\paragraph{Spiking Music compression quality}
The FREE Spiking Music Compression yielded a high reconstruction quality which is similar to the RVQ baseline at the same compression bit-rate. In terms of average SI-SNR, the reconstruction was better with RVQ $11\pm2$ for RVQ than with the dense model ($9\pm2$), however on the perceptual rating, the DENSE model was preferred over RVQ. The perceptual results are reported in Figure \ref{fig:exp}. Over $130$ listening tests, the FREE Spiking Music Compression was more frequently rated as better than RVQ samples rather than the opposite. Their average MUSHRA score was not significantly different.

\begin{figure*}[ht]
    \centering
    \includegraphics[width = 0.95\textwidth]{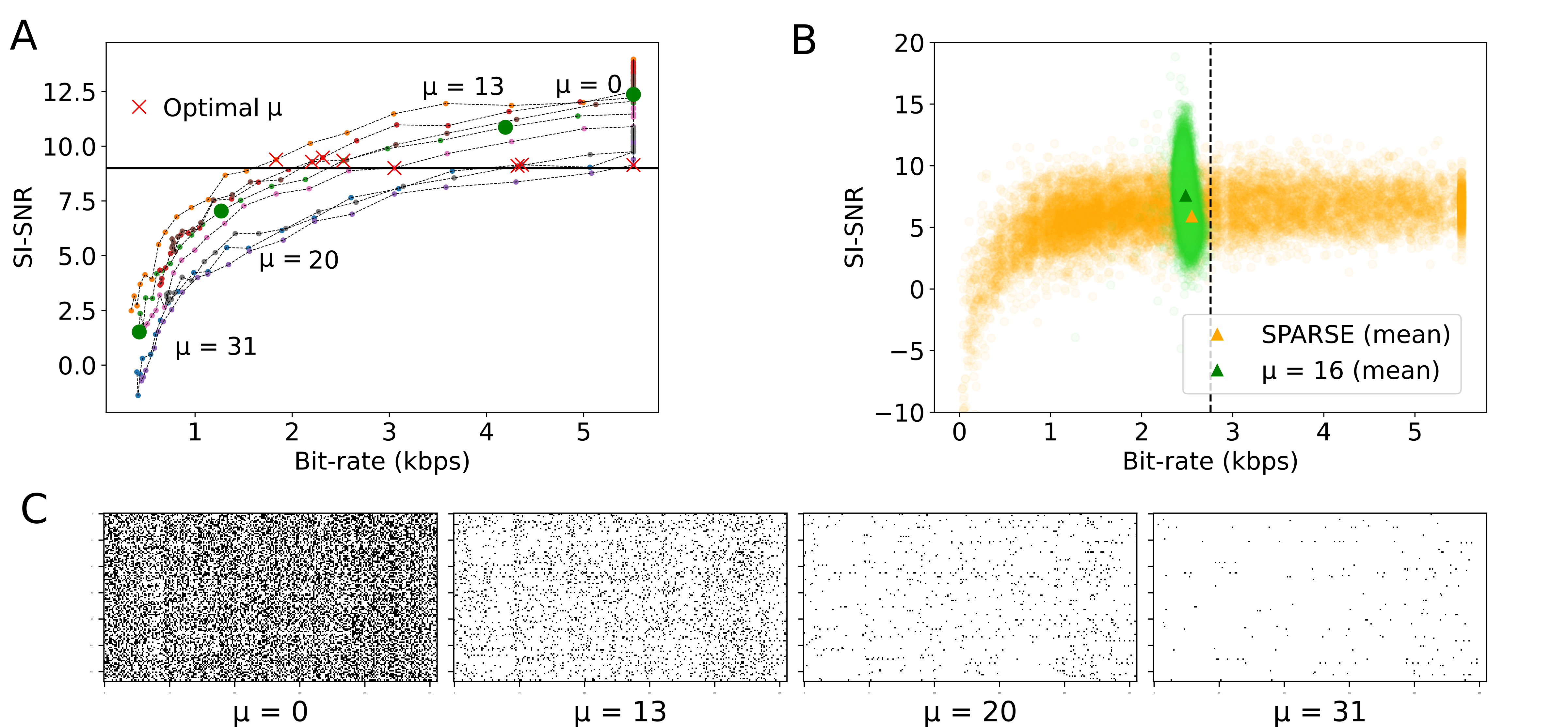}
    \caption{\textbf{Controllable reconstruction quality.} \textbf{A} Bit rate versus signal-to-noise ratio of 8 samples across the different compression levels $\mu$. The "quality-controlled" value of $\mu$ is the highest possible $\mu$ that reconstructs the sample with SI-SNR>9. The green balls are linked to the spiking representations in panel C. \textbf{B} Distribution of Bit rate versus SI-SNR ratio for the SPARSE and $\mu$-SPARSE with $\mu = 16$. The mean value is indicated by the triangle. Notice that the $\mu$ model ensures concentration around the target bit rate and limits the drops in SI-SNR. \textbf{C} Event matrix of the single audio sample at different compression levels.}
    \label{fig:mu_model}
\end{figure*}

\paragraph{Achieving low event rate without collapse}
When activating the representation loss to favor sparsity, the SPARSE spiking compression with sparse matrix representation format brought the resulting bit-rate below the baseline leading to an average bit-rate of $2.59$kbps. 
Achieving a functional sparse model was uneasy given that the model was susceptible to representation collapse: the audio reconstruction collapsed into pure noise and never generated good audio samples when we kept $\gamma=\gamma_\infty$ constant or with excessively high value $\gamma_\infty$.
To succeed we split the $1$M training steps of the SPARSE into three phases.
In the first phase of training, the sparsity loss coefficient was set $\gamma=0$ for long enough to generate recognizable audio similarly as in the FREE model. This model had therefore around $50\%$ zeros at the end of the first phase. In a second phase, $\gamma$ increased to bring the proportion of non-zero entries in the range of $10\%$ where the sparse matrix format becomes beneficial (see Figure \ref{fig:sparsity}). In this last phase, we keep $\gamma=\gamma_\infty$ to stabilize the representation.

Under this schedule, it became possible that audio reconstructed from the sparse binary matrices had satisfactory quality: all piano key strikes were clear and intelligible although there was a perceivable background noise. Although the reconstruction quality of the SPARSE model was worse than with the RVQ and FREE models, the quality was judged similar for audio samples reconstructed directly from a MEL Spectrogram. In our mind this is already a strong baseline since it corresponds to the default implementation of our diffusion decoder \cite{preechakul2022diffusion}.
This is visible in the MUSHRA online test where samples from the SPARSE model were judged as frequently better or worse than the MEL samples. 
To provide worse reference points, we also added an impaired RVQ model with $50k$ gradient descent time steps instead of $1M$ which achieved the lowest MUSHRA score. The anchor which was audio distorted by hand was rated even lower.

\paragraph{Controllable reconstruction quality}
Intuitively, some piano recordings have samples with high note density and others with sustained silenced. Interestingly we observed that the SPARSE Spiking compression allocated more events (therefore bits) when notes are frequent, and speculated the ideal event-based compression technique should have a variable event rate that adapts to the complexity of the encoded audio. Having this in mind, we studied the failure modes of the SPARSE Spiking Music Compression and found that the algorithm achieved very low SI-SNR on audio samples where it allocated very few events (see the orange samples with low bit-rate and SI-SNR in Figure \ref{fig:model_simple}B), we therefore explored a control mechanism to guarantee a stable audio quality with Spiking Music compression.

To exhibit a simple solution to this problem, we designed the $\mu$-SPARSE model as a variant where the sequence-to-sequence layer before and after the spiking compression additionally receive an integer $\mu \in \{0,1,2,...,31\}$ as a prompt of the bit-rate compression target.
This prompt is fed into the model as a learnable embedding representation added to the encoder's output or the decoder's input.
The model is then trained to achieve a target event count $S_0(\mu)= N T 2^{-\mu/4}$, by replacing the representation loss from equation \eqref{eq:Lz} by a sample-specific loss function of the form:
\begin{equation}
\mathcal{L}_{z}(\mu) = |~S - S_0(\mu)~|~.
\end{equation}
Given the direct relationship between event-rate and bit-rate given by equation \eqref{eq:B_ctime}, the main difference is the replacement of the rectified linear function by the absolute value.

In Figure \ref{fig:mu_model}B we show at test time, that an intermediate event-rate level $\mu=16$ avoided the failure cases where too few events are allocated for a specific audio sample resulting in very poor reconstruction quality.
Additionally, in a sender-receiver paradigm, the sender can, in fact, choose a strategical compression level $\mu^*$ for each sample, and transmit the compression level integer $\mu^*$ along with the compressed message. 
In Figure \ref{fig:mu_model}A, we display the bit-rate obtained when $\mu^*$ is chosen as the lowest compression level achieving SI-SNR higher than $9$. In this way, this $\mu-$SPARSE model guarantees a lower bound on the reconstruction quality. Among the samples visible in Figure\ref{fig:mu_model}A, most of the quality-guaranteed audio samples could be transmitted with approximately $2$kbps and a minority of hard audio samples would require a higher but reasonable compression bit rate.

\section{Spiking units are selective to piano notes}
\label{sec:corr}
\begin{figure*}
    \centering
    \includegraphics[width=\textwidth]{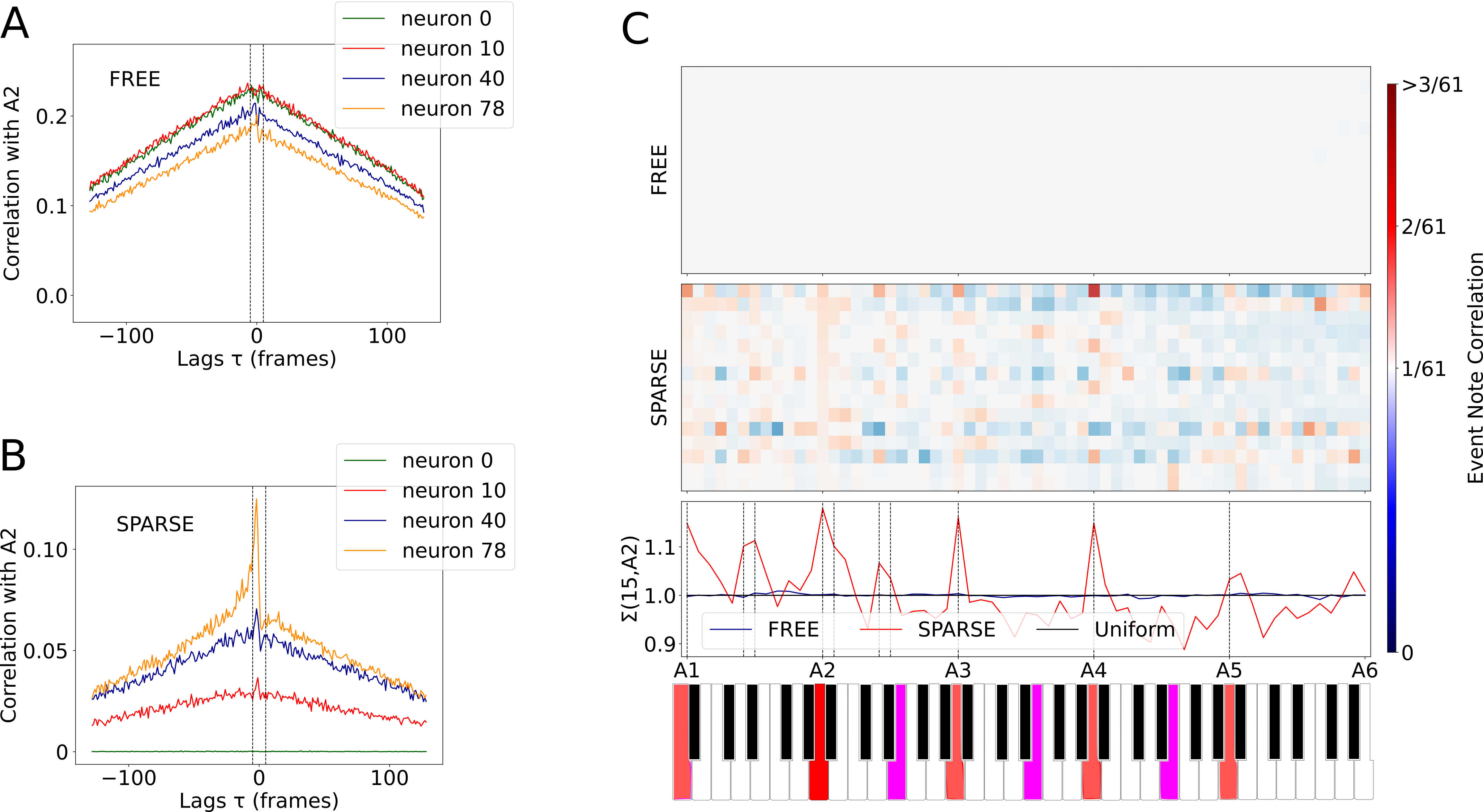}
    \caption{\textbf{Selectivity and synchrony with piano key strikes.} \textbf{A-B} Event-note correlation functions for FREE (\textbf{A}) and SPARSE (\textbf{B}) models. \textbf{C} We display the distribution of the peak prominence $\phi^{i\alpha}$ for all notes in the range A1-A6. We show the 15 units with the highest correlation with $\alpha=A2$. The bottom plot shows the distribution of peak prominence for the first unit.}
    \label{fig:midi_corr}
\end{figure*}

Since the MAESTRO dataset is constructed from MIDI piano music, the Spiking Compression algorithm could in principle re-discover the MIDI messages and encode key-stroke via as few spike events. We wondered therefore whether the event-base encoding discovered with Spiking Music compression relates to the keystrokes on the recorded piano and whether we could identify additional harmonic information like musical chords or tonality. To study this, we analyzed the cross-correlation between spiking units $z^i$ in the auto-encoder bottleneck and MIDI-note onsets of the note $\alpha$ represented as the binary event sequence $n^\alpha$ (exact MIDI-note onsets are provided in the MAESTRO dataset). This cross-correlation is defined below, with angle brackets denoting an average over the validation split of the MAESTRO dataset:
\begin{equation}
C^{i\alpha}(\tau) = \left<\sum_t z^i_t n^{\alpha}_{t+\tau}\right>~.
\end{equation}
To measure when the events of unit $i$ are selective and synchronized with the note $\alpha$ (e.g. the note $\alpha=A3$ or any other note of the piano keyboard), we introduce the peak prominence $\phi^{i\alpha}$ of the cross-correlation function as the normalized integral of the cross-correlation function $C^{i\alpha}(\tau)$ over short time lags smaller than $10$ time-steps ($\approx 0.23s$).
If the prominence is high $\phi^{i\alpha}$, it means that the unit $i$ is often spiking in synchrony when the note $\alpha$ is hit on the piano keyboard. Furthermore, the distribution of $\phi^{i\alpha}$ across all notes $\alpha$ shows how much the unit $i$ is carries harmonic content: for instance, if this distribution is uniform it means that the unit $i$ is not selective to any note $\alpha$.

Figure \ref{fig:midi_corr} summarizes the findings of our analysis.
We found a clear difference in the shapes of the cross-correlation functions between the SPARSE and FREE models. The sharp peaks at lag $\tau=0$ of the cross-correlation are characteristic of the SPARSE model seen in panel B. In panel C we show the peak prominence $\phi^{i\alpha}$ across $15$ spiking units and $88$ piano note pairs. We selected the $15$ units that correlate the most with the note $\alpha=A2$ in each model. For the FREE model, this matrix is consistently white which tells us that all $15$ units carry much less harmonic information than in the SPARSE model. In comparison to the SPARSE model, the darker shades of blue and red reflect very low or high peak prominence, indicating that spiking units are selective to specific note events. In the bottom plot in panel C, the peak prominence $\phi^{i\alpha}$ for the first note of the SPARSE model matrix is shown. It shows that the unit $i$ is tuned to multiple octaves of the note $A$ (highlighted in red). The same unit is also partially tuned to the note $E$ (pink) which is the fifth degree in the $A$ scale.  Interestingly, it is positively tuned to the notes $E1$ and $E2$ but negatively tuned to the other higher octaves of $E$.

\section{Discussion}
\label{discussion}

We have described Spiking Music compression as a method to compress audio recordings.
This method relies on a deep binary auto-encoder, so each audio sample can be reconstructed from the event matrix $\boldsymbol{z}$.
We have described two variants of the model: In the FREE model without sparsity loss, the reconstruction quality is competitive with the RVQ-VAE which is the state-of-the-art for audio compression \cite{defossez2022high}. This is surprising because the RVQ baseline requires a finely tuned k-means algorithm to learn the codebook which is not needed with our binary end-to-end model. The second Spiking Music compression variant includes an auxiliary sparsity loss. With a sparsity loss schedule, it becomes sparse enough to enjoy storage benefits with a sparse matrix storage algorithm. In this regime, the model spontaneously exhibits selectivity and synchrony with piano key strike events.

\section*{Impact statement}
This paper presents work whose goal is to advance the field of machine learning. There are many potential social consequences of our work, we would like to highlight that sparsity and event-based computation is thought to be a promising direction towards energy-efficient computing hardware.

\section*{Acknowledgments}
We would like to thank Martin Barry for helping with the crowd-sourcing platform, and Christos Sourmpis, Wulfram Gerstner, and Vivien Seguy for helpful discussions. This research was supported by the Swiss National Science Foundation with grant No. $200020\_207426$, the Sinergia grant No. $CRSII5 198612$, and an Intel INRC research grant.

\bibliography{references}
\bibliographystyle{abbrvnat}

\end{document}